\documentclass[preprint, su perscriptaddress, nonumbers]{revtex4}
\pdfoutput=1 
\usepackage{bm}
\usepackage{graphicx}
\usepackage{color}

\begin{document}

\title{  III-nitride tunable cup-cavities supporting quasi whispering gallery modes from ultraviolet to near infrared}

\author{T. V. Shubina}
\email{shubina@beam.ioffe.ru}
\affiliation{Ioffe Institute, 26 Polytekhnicheskaya, St. Petersburg 194021, Russia}
\author{G.~Pozina}
\affiliation{Department of Physics, Chemistry and Biology (IFM), Link\"oping University, S-581 83 Link\"oping, Sweden}
\author{V.~N.~Jmerik}
\author{V.~Yu.~Davydov}
\affiliation{Ioffe Institute, 26 Polytekhnicheskaya, St. Petersburg 194021, Russia}
\author{C.~Hemmingsson}
\affiliation{Department of Physics, Chemistry and Biology (IFM), Link\"oping University, S-581 83 Link\"oping, Sweden}
\author{A.~V.~Andrianov}
\author{D.~R.~Kazanov} 
\affiliation{Ioffe Institute, 26 Polytekhnicheskaya, St. Petersburg 194021, Russia}
\author{S.~V.~Ivanov}
\affiliation{Ioffe Institute, 26 Polytekhnicheskaya, St. Petersburg 194021, Russia}

\begin{abstract}

{\bf  

Rapidly developing nanophotonics needs microresonators for different spectral ranges, formed by chip-compatible technologies. In addition, the tunable ones are in greatest demand. Here, we present epitaxial site--controlled III--nitride cup--cavities which can operate from ultraviolet to near--infrared, supporting quasi whispering gallery modes up to room temperature. In these cavities, the refractive index variation near an absorption edge causes the remarkable effect of mode switching, which is accompanied by the change of spatial intensity distribution, concentration of light efficiently into a subwavelength volume, and  emission of terahertz photons. At a distance from the edge, the mode-related narrow emission lines have stable energies and widths at different temperatures. Moreover, their energies  are identical in the large 'ripened' monocrystal cavities. Our results shed light on the mode behavior in the semiconductor cavities and open the way for single--growth--run manufacturing the devices comprising an active region and a cavity with tunable mode frequencies.}

\end{abstract}

\maketitle

Whispering gallery modes (WGM) microresonators have attracted growing attention as essential building blocks of nano emitters$^{1-3}$, detectors, biological and chemical sensors$^{4-6}$. They are especially advantageous for structures made from wide--gap III--nitrides$^{7-11}$, because manufacturing the planar Bragg microcavities is difficult with decreasing the thicknesses of constituent layers in the ultraviolet (UV) range. Since a wavelength determines the characteristic size of the WGM microresonator as a whole, a similar problem for them will arise  at shorter wavelengths.  On the other end of the spectral range, the narrow-gap III--nitrides could cover the near infrared (IR) range corresponding to the minimal losses in communication networks. However, the quality ($Q$) factor of the WGM resonators reads as $Q = E / \Delta E$, where $E$ is the energy and $\Delta E$ is the width of  WGM-related narrow lines. Consequently, the $Q$--factor should go down in the narrow--gap structures. Thus, it is a challenge to design the WGM cavities from the III--nitrides which, in general, possess good prospects for nanotechnology.

The narrow lines, which appear in the spectra of WGM structures,  are commonly associated with the emission enhancement due to the Purcell effect$^{12}$, when spontaneous emission probability at the resonance wavelength $\lambda$ is increased  by a factor of $F_{P} \propto Q\lambda^{3}/n^{3}V$, where $n$ is the refractive index of the resonator and $V$ is the effective volume of an optical mode. This effect controls also the mode scattering into light$^{13}$. From the other hand,  as demonstrated by Kleppner$^{14}$, the inhibition of spontaneous decay rate by a factor of order $F_{P}$ can take place  when the density of final photon states or cavity size are too small,  or else with detuning from the resonance$^{15}$. We assume that the mode intensity distribution in a cavity, which in principle is spatially inhomogeneous, should control  the local performance of these  effects.  However, not much attention was paid to that with the investigations of the intensity pattern in the WGM cavities$^{16,17}$. 

Currently, there is searching the cavity design which satisfies three basic requirements: i) high $Q$-factor, ii) tunability of working frequencies,  and iii) compatibility with the chip technology. It is worth mentioning that the $Q$--factor can hardly be very high in "open" cavities which emit light. However, it does not present a serious obstacle, since the resonators of the moderate quality  might be quite satisfactory for many nanophotonics applications. To tune WGMs in the solid cavities, there are few methods: temperature$^{18}$, strain $^{19}$, and shape variation, like in the so-called "bottle microresonators"$^{20}$. In particular, the temperature-induced shift at the optical pumping approached 3.7 meV in a hybrid structure comprising a microsphere and  a II-VI nanoparticle $^{21}$. 

In many cases, the WGM cavities are formed by post--growth etching$^{7-9}$; alternatively, they can be created by epitaxy$^{16,22,23}$. Both approaches possess inherent advantages and disadvantages. The etching allows fabricating the structures of large diameters, which can maintain the high order modes corresponding to the lower decay rate and enhanced $Q$--factor. However, the number of modes existing in the vicinity of the principal one increases$^{24}$, that complicates spectral selection of discrete transitions. The rotational symmetry of the etched discs prohibits coupling the free--space light into and from such a cavity, especially if it possesses the high $Q$--factor, when the WGMs are strongly confined inside. This leads to the complication of optical schemes by introducing either elements  for  coupling$^{3}$ or special scatters$^{25}$. Epitaxial techniques, in general, offer a nice opportunity to fabricate a cavity and an active region during the same growth run. These cavities are nano-- and microcrystals, whose precise location on the growth surface can be organized$^{26}$. The  sizes of the epitaxial cavities are more suitable for supporting the low-order modes$^{27}$. On the other hand, the grown objects have usually a shape which reproduces the crystal structure, i.e. hexagonal in III--nitrides and II--oxides. Distortion of this shape  may unpredictably change the WGM frequencies$^{22}$.  

Here, we demonstrate for the first time monocrystal cup--cavities with reproducible sizes and frequencies, fabricated by  molecular beam epitaxy (MBE) from III--nitrides. GaN and InN were chosen to prove the concept of the cup--cavities for the wide spectral range from UV to near IR.  The path of the light within these cavities is 'polygonal' in nature; therefore, the cavity modes are called as quasi--WGMs. Their frequencies and intensity distribution deviate from those in a disc, and a characteristic size can be smaller for the same wavelength. The tunability of the quasi--WGMs in the cup--cavity is conditioned by strong optical dispersion in the semiconductors near an absorption edge,  which can be controlled not only by temperature but also carrier density variation. We report on the mode switching with the energy change of $\sim$14--meV,  which is accompanied by the modification of spatial intensity distribution and emission of terahertz (THz) quanta. In some degree, the cup--cavity operates as a parabolic mirror concentrating the light.  At room temperature, the dominant radial type of the quasi--WGMs  provides such a concentration in a small $\sim$300--nm--sized area in the central part of the cup--cavity, that can be used for selective enhancement of a limited number of quantum emitters.    
\\
\\
{\bf RESULTS} 

{\bf Samples  }

Our cup--cavities were grown on cone-shaped patterned substrates by MBE which has been developed to  create  the site--controlled monocrystals of different shapes, such as cups and nanocolumns (Fig. 1). The latter are promising for single--photon emitters$\bf ^{26}$. Details on the MBE technology are given in the Methods and Supplementary Fig. 1. Here, we claim that  the metal--rich conditions are preferable for the cup--cavity formation. The creation of the InN cup--cavity was a challenge, since InN is the most problematic  material for application among III--nitrides. Importantly, the wider PL band  in InN is very suitable to depict the almost full set of the anticipated modes to study the switching effect.  This would be impossible using the GaN crystals, where the PL line is typically narrow and can coincide only with one mode (Fig. 1 (c)). Therefore, most of the results presented below concern the InN cup--cavities.

\begin{figure}%
\includegraphics{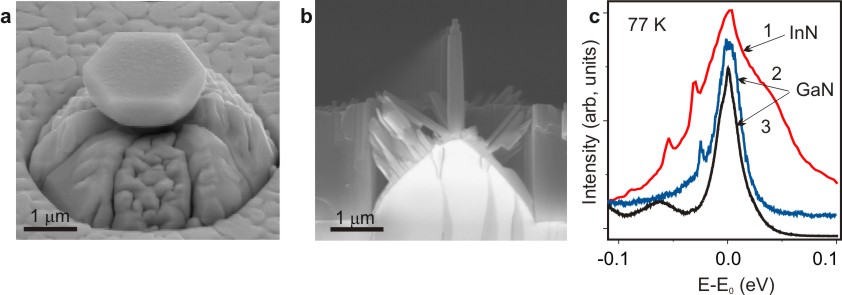}
\caption{\label{fig1} Typical SEM images and $\mu$--PL spectra of monocrystals grown by MBE on cone-shaped patterned substrates. (a) A cup--cavity,  (b) a nanocolumn array shown from the cleaved facet of a sample. (c) $\mu$--PL spectra measured in: 1) the InN cup--cavity,  2) the GaN cup--cavity, and 3) the GaN nanocolumns. The spectra are  normalized and arbitrary shifted for the sake of demonstrativeness; the energy counts off from the PL peak energy E$_{0}$ which is taken as 0.758 eV  for InN and 3.472 eV for GaN. The narrow lines are absent in the nanocolumn spectrum because their small diameter  does not allow supporting the WGM modes.   }
\end{figure}

According to scanning electron microscopy (SEM) studies, a significant part ($\sim$20$\%$) of the monocrystals  possesses the ideal shape and large characteristic diameter ({\em d}). Such a yield is sufficient for practical applications; moreover, it can be further improved. Currently, the maximal {\em d}  achieved for the InN cup--cavities ($\sim$2.2 $\mu$m) exceeds that for the GaN ones ($\sim$1.2 $\mu$m). However, in both cases the low-order optical modes can be supported, since the light wavelength inside GaN ($\sim$0.12 $\mu$m) is much smaller than that in InN ($\sim$0.55 $\mu$m).  

\begin{figure}[t]
\includegraphics{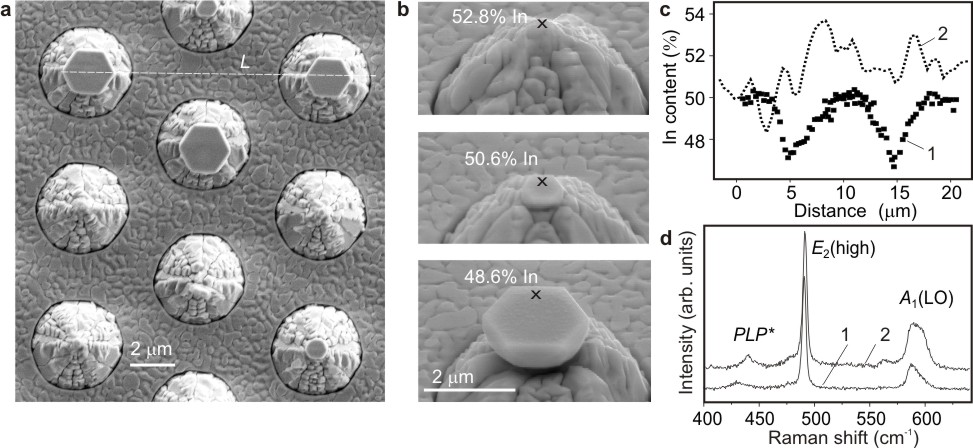}
\caption{\label{fig2} Improved structural properties of InN monocrystals as compared with planar layers. (a) A plane--view SEM image showing smoother top surface of the crystals than the surrounding planar area. (b) Evolution of the crystal size and composition during growth. The points of EDX microanalysis are marked by '\textsf{x}'; the In content is given nearby.  (c) The EDX profiles of In content done: 1$-$along a line $\it{L}$ in (a); 2$-$across the area enriched with metallic indium in a reference planar layer.  (d) First-order $\mu$--R spectra taken from: 1$-$the crystal, 2$-$the planar layer. The crystals exhibit the narrower widths of allowed Raman modes, E$_2$ and A$_1$(LO), and the shifted towards lower frequencies mixed plasmon--LO--phonon mode PLP$^{*}$. }
\end{figure}

\begin{figure}[t]
\includegraphics{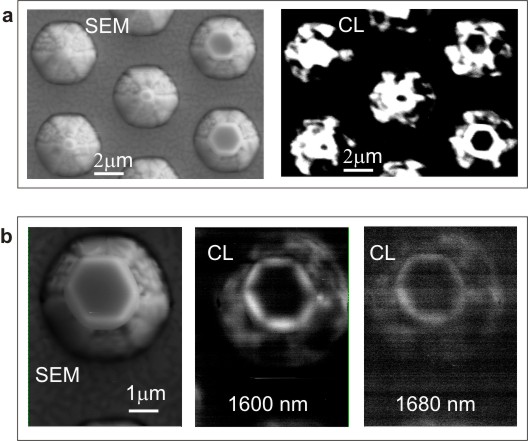}
\caption{\label{fig3} SEM and $\mu$--CL images recorded at 5 K  from an InN sample with the cup-cavities using different CL regimes. (a) SEM and panchromatic CL images of the same area with the cup-cavities of different sizes.  (b) SEM and mono--CL images recorded  from the same single  crystal with the central  wavelengths  of detection: 1600 nm (0.775 eV), and 1680 nm (0.738 eV). }
\end{figure}

The monocrystals grown on the patterned substrate exhibit better structural quality than a reference layer grown using the identical MBE regimes on a planar substrate. It has been confirmed by SEM,  energy dispersive x--ray (EDX) microanalysis, and micro--Raman ($\mu$--R) studies (Fig. 2). The crystal surface is smoother that that of a surrounding area. In accordance with previous results$\bf ^{28}$, the Raman data correspond to the lower densities of structural defects and decreased concentration of free carriers. In general, the InN  tends to non--stoichiometry and spontaneous formation of metallic In nanoparticles$\bf ^{29}$. Plasmonic resonances in  such nanoparticles could enhance emission in a semiconductor matrix nearby or induce additional ohmic losses, depending on their sizes$\bf ^{30}$. When  considering the amplification of the emission intensity in our cup--cavities, we should exclude the plasmonic enhancement as a possible mechanism, because no signs of the In excess  were found in the large InN crystals. 

Micro--cathodoluminescence ($\mu$--CL) studies showed that the brightest emission spots are always at the crystals (Fig. 3 (a)) that is consistent with their high material quality. However, this emission is not homogeneous but reflects the spatial distribution of electromagnetic energy within the cavities. It is interesting that at low temperature the smaller crystal size the higher emission intensity; obviously, the leaky modes prevail in these low--quality cavities. 
In the mono--CL images (Fig. 3 (b)), the brightest emission patterns were recorded at the energies of the strongest narrow lines observed in the micro--photoluminescence ($\mu$--PL) spectra (Fig. 4). At the high--energy side of these spectra, where the narrow lines are absent, the emission from the crystals merges with the background emission from the surroundings. 

When the temperature increases, the emission from the small crystals quenches fast due to their low cavity quality, while in the large ones it can be detected up to room temperature (Fig. 5 (a)). The decay time of emission is about 150 ps at low temperature (10 K). This time is noticeably shortened with the temperature rise, being $\sim$80 ps at 300 K. Because the emission from the small crystals is quenched, one can consider this decreased time as a characteristic of the large crystals themselves. In this case, it can be  caused partly by the acceleration of the recombination rate due to the Purcell effect in these cup-cavities.
\\
\\
{\bf   Quasi whispering gallery modes }

\begin{figure}[b]
\includegraphics{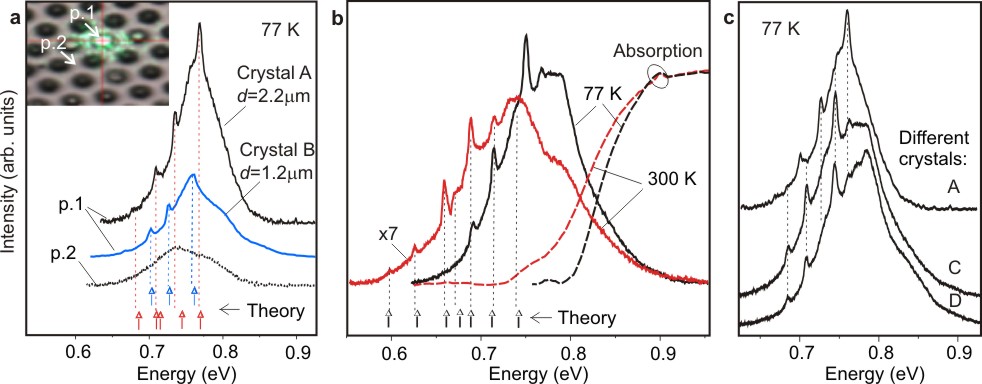}
\caption{\label{fig4} WGM--related narrow lines in $\mu$--PL spectra. (a) The inset demonstrates the spot of a laser beam (532 nm), which can be focused either on a crystal (p. 1) or on the area between the cones (p. 2). No narrow lines were observed in the latter case. The energies of the narrow lines are not identical in the crystals of different sizes: A ($d=$2.2 $\mu$m) and B ($d=$1.2 $\mu$m).  (b) $\mu$--PL spectra of the crystal C ($d=$2.2 $\mu$m)  measured at 77 K and 300 K, shown together with integral absorption spectra (dashed lines). The narrow line energies at a distance from the absorption edge are identical at different temperatures. Theoretical energies are marked at the bottom axes in (a) and (b).  (c) $\mu$--PL spectra measured in different crystals with $d\simeq$2.2 $\mu$m, exhibiting the similar energies of the WGM--related narrow lines.  }
\end{figure}

The spectra  of $\mu$--PL  measured in the large InN crystals comprise narrow WGM--related lines superimposed on a wide emission band. Such lines are absent with measurements from the planar area (Fig. 4 (a)). The distinctiveness and strength of these narrow emission lines in the cup--cavities are comparable with those in the disc resonators fabricated by etching$\bf ^{7,9}$. The $Q$--factor is estimated as $\sim$700 in the GaN  cavities, being $\sim$200 in the InN ones (the decreased value is mainly because InN has  smaller emission energy). Spectral separation between the WGMs varies from 13 meV to 30 meV that  permits the selective enhancement of optical transitions.

In the cup--cavities, the narrow emission lines are observed at the maximum and lower-energy part of the emission spectra. At the higher-energy side they are suppressed, mainly because of optical losses induced by a close  absorption edge.  With a temperature rise, the broadened absorption edge quenches those of the modes, which were at the maximum. Instead, the new modes come into play at the lower--energy side of the PL band when it shifts following the edge. Between these boundaries, a set of the narrow lines can be found, whose positions are stable with increasing the temperature from 77 K to 300 K (Fig. 4 (b)). It is essential that the narrow lines  have insignificant broadening  in such a wide temperature range. 

The amazing finding is the observed identity  of the mode frequencies recorded in all 'ripened' crystals of the largest size, which were chosen from the different parts of a wafer (Fig. 4 (c)). This identity, existing within the limits of experimental accuracy, means that variation of shape and characteristic sizes of the large microcrystals is negligible. Such a similarity implies that the self--limitation (saturation) of the crystal sizes takes place at the prolonged  MBE growth. Note that the standardized mode frequencies are a prerequisite for any cavity application.
\\
\\
{\bf  Mode switching}

The remarkable effect of mode switching was discovered by  the $\mu$--CL measurements performed in the 5--300 K range. At low temperatures, the spatial distribution of the $\mu$--CL signal corresponds to the azimuthal type, when the highest density of electromagnetic field is at the periphery of the crystal. Such an emission pattern can be observed even in the rather small crystals of $\sim$1 $\mu$m diameters (Fig. 3 (a)).   When the temperature increases up to 100$-$150 K, the emission pattern starts to change. At room temperature, it corresponds to the radial type when the most bright spot occurs in the center (Fig. 5 (a)).

To shed light on this phenomenon, let us remind  that  in the simplest cylindrical approximation the frequency of a resonator mode of the $m$--order  obeys the law $\omega\propto mc/dn $, where $d$ -- diameter, $n$ --  refractive index, $c$ -- speed of light. Therefore, with the constant geometry, i.e. when the same crystal is measured and  the small thermal expansion of its size can be neglected, only variation of $n$  will change the type of the modes with increasing the temperature. The variation of the absorption coefficient will rather suppress the modes, as it was observed at the higher--energy side of the $\mu$--PL spectra.  

\begin{figure}[b]
\includegraphics{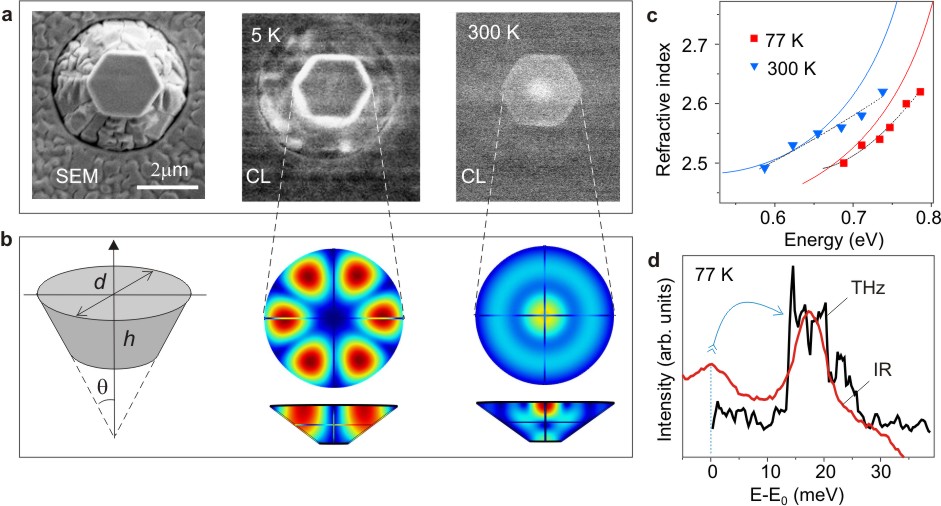}
\caption{\label{fig5} Temperature--induced mode switching in the cup-cavities. (a) SEM and $\mu$-CL images of the same InN crystal recorded at 5 K and 300 K showing the change of the dominant mode type from azimuthal to radial. (b) The simulated distributions of electromagnetic field intensity, as seen in the plane and cross--section of the crystal, shown together with the schematic shape used in these simulations. (c) Variation of the refractive index $n$ derived from the fitting of the WGMs at low (squares) and room (triangles) temperatures; the dotted lines is their polynomial fitting; the solid lines indicate the $n$ dependencies derived from the integral optical spectra. (d) Teraherz spectrum shown together with the peak of the IR emission comprising two unstable WGM--related lines. The energy scale is taken as $E-E_{0}$, where  $E_{0}$  is the peak energy of the first line for the IR emission, while for the THz band $E_{0}=0$.}
\end{figure}

We have performed the numerical simulation of the quasi--WGMs in the cup--cavities to prove the  anticipated effect of the $n$ variation on the mode switching. The examples of the spatial distributions of the electromagnetic field intensity inside the cavities at different temperatures are depicted in Fig. 5 (b). The extended sets of simulated patterns of mode intensity distribution for both InN and GaN cavities are given in the Supplementary Fig. 2. 

To determine the $\textit{n}$ dispersion, needed for these simulations, we have measured first the integral absorption and reflection; then, the complex dielectric function was extracted by applying the Kramers--Kronig relations. The obtained parameters were reasonably consistent with published dependencies$\bf ^{31,32}$. 
More accurate analysis of frequencies and energy distributions was done using $\textit{n}$  as a fitting parameter. Namely, its value was varied  to find a feasible mode in the vicinity of the experimentally observed narrow line. 

Figure 5 (c) shows the refraction index values derived from such simulations. It deviate somewhat from the dependencies obtained by analysis of the integral optical spectra.  This difference arose because an effective absorption edge in an InN layer can shift towards lower energies due to the optical losses induced by the metallic In nanoparticles and defect states, while the shift in the opposite direction may be when the material is In--depleted$\bf ^{29}$. Bearing that in mind, we can conclude that the modelling of the modes represents a unique way to determine the optical parameters characteristic for the crystal material itself.

The modelling has confirmed that the azimuthal type of quasi--WGMs prevails over the radial type at low temperatures. This situation inversely changes at room temperature.  For the GaN cavities, the modelling has shown  that the number of the modes within the emission band is limited: only three distinct modes were found. The theoretical mode at 3.450 eV  coincides reasonably with the narrow line at 3.447 eV in the experimental spectrum shown in Fig. 1. We assume that the two other modes, anticipated near 3.47 eV, can be suppressed by neutral--donor bound exciton scattering, which provides the strong attenuation of propagating light in GaN$\bf ^{33}$ 

The switching between two neighboring modes has to gain the small energy excess, which can radiate as the quanta of terahertz (THz) emission. This suggestion is consistent with a THz spectrum shown in Fig. 5 (d), which was measured using electric-pulse excitation technique$\bf ^{34}$. Apparently, the THz peak energy at 14 meV ($\sim$3.5 THz) corresponds well to the energy separation (13--15 meV) between the adjacent modes at the maximum of the IR emission band.  Note that the modes are absent at the higher-energy side of the band and that their separation increases up to 20--30 meV at the low--energy wing. Such a situation may be a reason of the strongly asymmetrical shape of this THz spectrum. 
\\
\\
{\bf Discussion}

We highlight that the  refractive index variation, controlling the WGMs, is achievable in the semiconductor cavities not only by the temperature increase, but also by laser beam pumping and  electric-pulse propagation. In contrast with the slow temperature exposure, their  heating effect is  fast. In addition, these methods can change the concentration of free carriers,  which in turn can affect the complex refractive index to a considerable degree. In a degenerate semiconductor like InN, the Burstein--Moss effect will push the apparent absorption edge towards higher energies due to the filling of conduction band states. Therefore, its action will be opposite to the temperature--induced red shift of the absorption edge. 

Figure 6 shows the energy dependencies on the refractive index calculated for two neighboring modes, assuming that the carrier concentration is increased by either optical or electrical pumping.  For the realistic range of $(1-5)\cdot10^{12}$ cm$^{-2}$, the modification of the refractive index comprises several percents from an initial value; the imaginary part exhibits noticeable reduction as well$^{35}$. In this case, two scenarios can be realized depending on the applied power and exposure duration. First, the mode energy  can be increased by several meV with the conservation of the spatial intensity distribution. Such smooth tuning can be used to match the mode frequency with a resonance line in the sensors or nano--emitters. Alternatively, the abrupt switching to the neighboring mode can take place with emitting the extra energy as THz quanta. This effect could be promissing for the optical switchers. 

The weighty argument in favor of the mode switching is the emitting of the THz photons, whose energy is equal to the energy gap between the adjusting modes. Since the used regime with pulsed excitation prevents overheating the samples, we assume that the increase of the carrier concentration governs the process. Note that the THz emission in InN could arise due to the surface plasmons coupled to electromagnetic field at the Bragg grating which is formed by the cones on the patterned substrate. However, our estimations showed that for the used cone period the emission frequency should be below 2 THz, while the signal from the InN structure with the monocrystals was registered at $\sim$ 3.5 THz.  

\begin{figure}[t]
\includegraphics{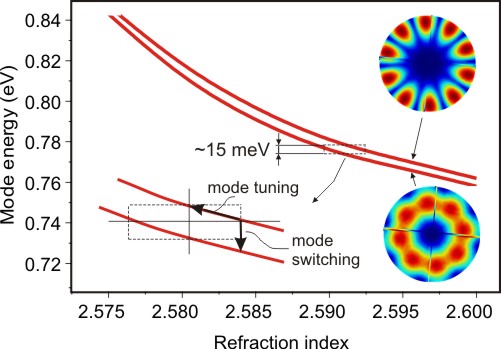}
\caption{\label{fig6} Dependencies of mode energy on the refractive index calculated for two neighboring modes in the InN cup--cavities. The simulated  spatial distributions of electromagnetic energy for these two modes are presented. The inset illustrates two possible regime: tuning and switching. 
}
\end{figure}

Considering the emission intensity pattern in the low--temperature $\mu-$CL images we should draw attention to the strong contrast between the emitting periphery  and dark central area, which reflects the emission dynamics in the weak coupling regime within these areas.  Neglecting the decay into leaky modes and possible detunibg, the ratio of spontaneous emission rate $1/\tau$ of a radiating dipole located at position $r$ in the cavity relatively to the rate in a homogeneous medium $1/\tau_0$ can be expressed as ${{\tau_{0}} / {\tau}} \propto F_{P}({{|E(\textbf{r})|^{2}} /  {|E_{max}|^{2}}})$, where  the electric field  distribution in the cavity $E(\textbf{r})$ has a maximum amplitude $E_{max}$$^{36}$. In the local areas where the density of photon states given by $E(\textbf{r})$ is close to zero, the emission is suppressed, although the overall Purcell factor $F_P$ can be high. 
At room temperature, the mode energy  is concentrated in a small area of a $\sim$300 nm diameter (Fig. 5 (a)), while in the rest area the emission is strongly inhibited. Such a phenomenon can be used  to select the limited number of radiating dipoles in a semiconductor cup-cavity instead of mesa formation, e.g., by expensive electron-beam lithograthy.

In summary, we demonstrate novel cup--cavities operating from UV to near--IR, which are fabricated solely by MBE from III--nitrides $-$ GaN and InN. Realization of these cup--cavities evidences the  potential of MBE for the cavity nanotechnology. Note that up to now there is no devices based on InN, while this semiconductor has been intensively studied since the beginning of this century. Therefore, fabrication of the InN cup--cavities  with stable mode frequencies is certain technological breakthrough in this field. The fabricated cup--cavities support the quasi--WGMs which are well--separated in frequency, that enables selective interaction with discrete quantum transitions. The cavity $Q$--factor, approaching 700 in the UV range, is sufficient for enhancement of the nano--emitter efficiency and detection sensitivity. Furthermore, we show that the WGMs in a semiconductor cavity can be controlled by the complex refractive index variation near an absorption edge by temperature or carrier concentration. The modes will have either stable or tunable frequencies depending on their position with respect to this edge. For the modes at the maximum of an emission band, it is possible to realize both the mode tuning with the conservation of the spatial intensity distribution and switching between adjusting modes with emitting the THz photons. In general, our results open a way to realize the semiconductor WGM cavities with tunable frequencies controlled by optical pumping or electric-pulse applying. 
\\
\\
{\bf METHODS}
\\
\\
{\bf Molecular beam epitaxy growth} 

\begin{table}[b]
\caption{ \label{t1} MBE technology of InN and GaN monocrystals.  For the buffer growth, either 
the ordinary MBE or migration enhanced epitaxy (MEE) was used. } 
\begin{ruledtabular}
\begin{tabular}{llll}
Structure & Buffer & Buffer  & Basic  \\
 &   & growth parameters & growth parameters \\

\hline
InN cup-cavities  & 200-nm-thick &  MBE, $F_{Ga}/F_{N}\simeq1.6$,  & MBE, $F_{In}/F_{N}\simeq1.4$\\
                & GaN          &  $T_{s}=700^{\circ}$C              & $T_{s}=470^{\circ}$C   \\
\hline
GaN cup-cavities  & 70-nm-thick &  MEE, $F_{Al}/F_{N}\simeq1$,     & MBE, $F_{Ga}/F_{N}\simeq1.3$\\
                & AlN          &  $T_{s}=784^{\circ}$C           & $T_{s}=650^{\circ}$C   \\
\hline
GaN nanocolumns  & 40-nm-thick &  MBE, $F_{Ga}/F_{N}\simeq1.2$,   & MBE, $F_{Ga}/F_{N}\simeq0.8$\\
                & GaN          &  $T_{s}=605^{\circ}$C              & $T_{s}=750^{\circ}$C   \\
\end{tabular}
\end{ruledtabular}
 \end{table}
III--nitride monocrystal cavities were grown by the low--temperature plasma--assisted (PA) MBE on cone-shaped patterned sapphire substrates using a Compact 21T (Riber SA) setup equipped with a nitrogen plasma activator HD25 (Oxford Applied Research). The substrates have the regularly positioned  cones with a diameter and height of 3 and 1.5 $\mu$m, respectively.  These cones act as pedestals for the grown monocrystals. The formation of the crystals on their tops is mainly induced by the difference in growth rates along basic crystallographic directions, stimulated by chosen technological parameters. The important factors which control the crystal shape are the substrate temperature $T_{s}$, fluxes ratio $F_{Ga,In}/F_{N}$, and polarity N or Ga(In) set by a buffer. Their variation can change the shape of the monocrystals from inverted hexagonal pyramides to nanocolumns.  The  procedure of the monocrystal fabrication comprises consecutive growth of the thin  buffer layer and basic micrometer--sized layer at different  T$_{\textrm{s}}$ and $F_{Ga,In}/F_{N}$.  The typical technological regimes are given in Table 1. The growth procedures are illustrated by SEM images given in Supplementary Fig. 1. 
\\
\\
{\bf Measurements} 

The $\mu$--PL studies were carried out in a cryostat at 77 and 300 K  with the spectral resolution of $\sim$0.5 meV under cw excitation by 532 nm and 325 nm laser lines for InN  and GaN structures, respectively. The beam impinging normally to the surface was focused by an objective into a spot with FWHM of $\sim$1 $\mu$m that is enough to measure separately the crystals. The same objective collected the PL signal and the sample image which is monitored by a charge--coupled detector (CCD). The PL was relayed to the slits of a monochromator and then to the nitrogen cooled CCD.  The $\mu$-R measurements  were done in a similar way using a specially designed Raman spectrometer integrated with an optical microscope. The time--resolved PL measurements were performed in a closed--cycle He cryostat using the near-IR-sensitive detector and a 710 nm line of a pulsed laser for excitation. The laser beam was focused into the 0.5--mm spot. The spectra of transmission and reflection were measured using a two-monochromators setup with excitation by a tungsten lamp focused into the 1-mm spot.  Both integral cw and time--resolved spectra characterize the properties of the areas which includes the crystals and surrounding layer. 

SEM, microanalysis, and $\mu$--CL studies were done using a LEO 1550 Gemini analytical scanning electron microscope equipped by a unit for EDX and a low-temperature $\mu$--CL stage with a nitrogen-cooled Ge--detector. At the EDX characterization performed with 0.5$\%$ accuracy, the nitrogen calibration was done using a perfect bulk GaN sample. The $\mu$--CL studies were carried out at 10 kV beam voltage from 5 K up to 300 K. Both panchromatic and mono--CL regimes were used. For the later, a CL signal was registered at particular wavelengths within the emission band.

The THz spectra were measured at 77~K with excitation by the series of packets of 
rectangular pulses with 15 V pulse height, 10 $\mu$s duration, and 71.5 Hz reputation rate. The pulse excitation prevents the heating of a sample. The pulses passed  through contacts formed on the top surface of a $~5$--mm--long sample. The step--scan Fourier spectrometer has the volume of optical path evacuated down to $6\ast10^{-2}$ Torr to exclude any influence of water vapor absorption on the shape of the THz spectra.  
\\
\\
{\bf Modelling} 

The  spatial distribution of the  electromagnetic field intensity inside the cavities  and mode frequencies were searched for the quasi--WGMs existing within the limits of emission bands of the cavity materials. It was done numerically by solving the Maxwell's equations using the Comsol Multiphysics software. We adopted boundary conditions so as to cause a rapid decay of light waves outside the crystal. The crystal dimensions were taken from the SEM data. In these simulations, the cup--like shape of the  cavities was approximated by a truncated cone. The prism-like shape was also used  (for GaN crystals). In this case, we revealed the distinct slab  (Fabry--P\'{e}rot) modes  propagating between two parallel facets.    
\\
\\
{\bf    Acknowledgments}                                               

The studies at the Ioffe Institute are supported by the Russian Science Foundation (Project $\#$ 14-22-00107). GP and CH are grateful for financial support from the Swedish Research Council (VR) and the Swedish Energy Agency. The authors thank D.~V.~Nechaev and N.~V.~Kuznetsova for growth assistance, T.~B.~Popova for material  characterization assistance, D.~I.~Kuritsyn, A.~N.~Smirnov and A.~O.~Zakhar$'$in for measurement assistance.  
\\
\\
{\bf Author contributions}

T.V.S. wrote the paper, coordinated optical studies and modelling; V.N.J. grown the samples; G.P. and V.Yu.D. performed optical studies; C.H. did material characterizations; A.V.A. provided THz measurements;  D.R.K. did  Comsol Multiphysics simulations, S.V.I. supervised the project. All authors participated in analyzing and discussing the data.
\\
\\
{\bf Competing financial interests}

The authors declare no competing financial interests
\\
\\
{\bf References}

1. Vahala, K. J. Optical Microcavities. {\em Nature} {\bf 424}, 839-846 (2003). 

2. Hill, M. T. $\&$ Gather C. Advances in small lasers. {\em Nature Photon.}  {\bf 8}, 908-918 (2014).

3. F\"{o}rtsch {\em et al.} A versatile source of single photons for quantum information processing. {\em Nature Commun.} {\bf  4}, 1818 (2013).

4. Vollmer, F. $\&$ Arnold S. Whispering-gallery-mode biosensing: label-free detection down to single molecules. {\em Nature Methods} {\bf 5}, 591 - 596 (2008).

5. Righini, G. C. {\em et al.} Whispering gallery mode microresonators: Fundamentals and applications. {\em Rivista Del Nuovo Cimento} {\bf 34}, 435-488 (2011).

6. Kippenberg, T. J. Particle sizing by mode splitting. {\em Nature Photon.}  {\bf 4}, 9-10 (2010).

7. Tamboli, A. C. {\em et al.} Room-temperature continuous-wave lasing in GaN/InGaN microdisks. {\em Nat. Photon.} {\bf  1}, 61-64 (2007).

8. Mexis, M. {\em et al.} High quality factor nitride-based optical cavities: microdisks with embedded GaN/Al(Ga)N quantum dots. {\em Optics Lett.} {\bf 15}, 2203-2205 (2011). 

9. B\"{u}rger, M. {\em et al.} Lasing properties of non-polar GaN quantum dots in cubic aluminum nitride microdisk cavities. {\em Appl. Phys. Lett.} {\bf  102}, 081105 (2013).

10. Baek, H., Hyun, J. R., Chung, K. $\&$ Oh, H. G-C. Selective excitation of Fabry-P\'{e}rot or whispering-gallery mode-type lasing in GaN microrods. {\em Appl. Phys. Lett.} {\bf 105}, 201108 (2014). 

11. Zhang, X., Cheung, Y. F., Zhang, Y. $\&$ Choi, H. W.  Whispering-gallery mode lasing from optically free-standing InGaN microdisks. {\em Opt. Lett.} {\bf 39}, 5614-7 (2014). 

12. Purcell, E. M. Spontaneous emission probabilities at radio frequencies. {\em Phys. Rev.} {\bf 69}, 681 (1946).

13. Kippenberg, T. J., Tchebotareva, A. L., Kalkman, J., Polman, A. $\&$  Vahala K. J. Purcell-Factor-Enhanced Scattering from Si Nanocrystals in an Optical Microcavity. {\em Phys. Rev. Lett.} {\bf 77}, 

14. Kleppner, D. Inhibited spontaneous emission. {\em Phys. Rev. Lett. } {\bf 47}, 106101 (1981).

15. Bayer, M. Reinecke, T. L., Weidner, F., Larionov, A., McDonald, A. $\&$ A. Forchel.  Inhibition and enhancement of the spontaneous emission of quantum dots in structured microresonators. {\em Phys. Rev. Lett.} {\bf 86}, 3168 (2001).

027406 (2009).

16. Kim, C., Kim, Y.-J., Yi, G.-C., $\&$ Kim, H. H. Whispering-gallery-modelike-enhanced emission from ZnO nanodisk. {\em Appl. Phys. Lett.} {\bf 88}, 88, 093104 (2006).

17. Mintairov, A. M. {\em et al.} High-spatial-resolution near-field photoluminescence and imaging of whispering-gallery modes in semiconductor microdisks with embedded quantum dots. {\em Phys. Rev. Lett.} {\bf 77}, 195322-7  (2008).

18. Vernooy, D. W., Furusawa, A., Georgiades, N. P., Ilchenko, V. S., $\&$  Kimble, H. J. Cavity QED with high-Q whispering gallery modes. {\em Phys. Rev. A} {\bf 57}, R2293-6 (1998).

19. Ilchenko V. S. {\em et al.} Strain-tunable high-Q optical microsphere resonator. {\em Opt. Commun.} {\bf 145} 86-90 (1998).

20. P\"{o}llinger, M., O'Shea, D.,  Warken, F. $\&$ Rauschenbeutel, A. Ultrahigh-Q Tunable Whispering-Gallery-Mode Microresonator{\em Phys. Rev. Lett.} {\bf 103}, 053901  (2009).

21. Grivas, C. {\em et al.} Single-mode tunable laser emission in the single-exciton regime from colloidal nanocrystals. {\em Nature Commun.} {\bf 4}, 2376 (2013).

22. Kuono, T.,  Sakai, M., Kishino, K. $\&$ Hara, K. Quasi-whispering gallery mode lasing action in an asymmetric hexagonal GaN microdisk. {\em Jap. J. Appl. Phys. } {\bf  52},  08JG03 (2013).

23. Tessarek, C. {\em et al.} Improving the optical properties of self-catalyzed GaN microrods toward whispering gallery mode lasing. {\em ACS Photon.} {\bf 1}, 990-997 (2014).

24. Kaliteevski, M.~A., Brand, S., Abram, R.~A., Kavokin, A. $\&$ Dang, L.~S. Whispering gallery polaritons in cylindrical cavities. {\em Phys. Rev. B} {\bf 75}, 233309 (2007).

25. Zhu, J. {\em et al.} Interfacing whispering-gallery microresonators and free space light with cavity enhanced Rayleigh scattering. {\em Scientific Reports} {\bf 4}, 6396-7 (2004).

26. Holmes, M.~J., Choi, K., Kako, S., Arita, M. $\&$ Arakawa, Y. Room-temperature triggered single photon emission from a III-nitride site-controlled nanowire quantum dot. {\em Nano Lett. } {\bf 14}, 982-986 (2014).

27. Nobis, T. $\&$ Grundmann, M. Low-order optical whispering-gallery modes in hexagonal nanocavities. {\em Phys. Rev. B} {\bf 72}, 063806 (2005).

28. Davydov V.~Yu. $\&$ Klochikhin, A.~A. Electronic and vibrational states in InN and In$_{x}$Ga$_{1-x}$N solid solutions. {\em Semiconductors } {\bf 38}, 861-898 (2004). 

29. Ivanov, S.~V., Shubina, T.~V., Komissarova, T.~A. $\&$ Jmerik, V.~N. Metastable nature of InN and In-rich InGaN alloy. {\em J. Cryst. Growth} {\bf  403}, 83-89 (2014).

30. Toropov, A. A. $\&$ Shubina, T.~V. {\em Plasmonic effects in metal-semiconductor nanostructures.}  (Oxford University Press, New York, 2015).

31. Goldhahn, R. {\em et al.} Dielectric Function of "Narrow" Band Gap InN. {\em Mat. Res. Soc. Symp. Proc.} {\bf 743} L5.9.1 (2003).

32. Watanabe, N., Kimoto, T. $\&$  Suda, J. The temperature dependence of the refractive indices of GaN and AlN from room temperature up to 515$\circ$C. {\em J. Appl. Phys. } {\bf  104},  106101-3 (2008). 

33. Shubina, T.~V. {\em et al.} B. Resonant Light Delay in GaN with Ballistic and Diffusive Propagation. {\em Phys. Rev. Lett.} {\bf 100},  087402 (2008).

34. Shubina, T. V. {\em et al.} Terahertz electroluminescence of surface plasmons from nanostructured InN layers. {\em Appl. Phys. Lett.} {\bf 96}, 183106 (2010). 

35. Bulutay, C., Turgut, C. M. $\&$ Zakhleniuk N. A. Carrier-induced refractive index change and optical absorption in wurtzite InN and GaN: Full-band approach. {\em Phys. Rev. B} {\bf 81}, 155206 - 7 (2010).

36. G\'{e}rard, J. M. {\em et al.} Enhanced spontaneous emission by quantum boxes in a monolithic optical microcavity, {\em Phys. Rev. Lett.} {\bf 81}, 1110-4  (1998).

-----------------
\\
\\
{\bf Figure Legends}
\\
\\
{\bf Figure 1}

Typical SEM images and $\mu$--PL spectra of monocrystals grown by MBE on cone-shaped patterned substrates. (a) A cup--cavity,  (b) a nanocolumn array shown from the cleaved facet of a sample. (c) $\mu$--PL spectra measured in: 1) the InN cup--cavity,  2) the GaN cup--cavity, and 3) the GaN nanocolumns. The spectra are  normalized and arbitrary shifted for the sake of demonstrativeness; the energy counts off from the PL peak energy E$_{0}$ which is taken as 0.758 eV  for InN and 3.472 eV for GaN. The narrow lines are absent in the nanocolumn spectrum because their small diameter  does not allow supporting the WGM modes.
\\
\\
{\bf Figure 2}

Improved structural properties of InN monocrystals as compared with planar layers. (a) A plane--view SEM image showing smoother top surface of the crystals than the surrounding planar area. (b) Evolution of the crystal size and composition during growth. The points of EDX microanalysis are marked by '\textsf{x}'; the In content is given nearby.  (c) The EDX profiles of In content done: 1$-$along a line $\it{L}$ in (a); 2$-$across the area enriched with metallic indium in a reference planar layer.  (d) First-order $\mu$--R spectra taken from: 1$-$the crystal, 2$-$the planar layer. The crystals exhibit the narrower widths of allowed Raman modes, E$_2$ and A$_1$(LO), and the shifted towards lower frequencies mixed plasmon--LO--phonon mode PLP$^{*}$.
\\
\\
{\bf Figure 3}

SEM and $\mu$--CL images recorded at 5 K  from an InN sample with the cup-cavities using different CL regimes. (a) SEM and panchromatic CL images of the same area with the cup-cavities of different sizes.  (b) SEM and mono--CL images recorded  from the same single  crystal with the central  wavelengths  of detection: 1600 nm (0.775 eV), and 1680 nm (0.738 eV).
\\
\\
{\bf Figure 4}

WGM--related narrow lines in $\mu$--PL spectra. (a) The inset demonstrates the spot of a laser beam (532 nm), which can be focused either on a crystal (p. 1) or on the area between the cones (p. 2). No narrow lines were observed in the latter case. The energies of the narrow lines are not identical in the crystals of different sizes: A ($d=$2.2 $\mu$m) and B ($d=$1.2 $\mu$m).  (b) $\mu$--PL spectra of the crystal C ($d=$2.2 $\mu$m)  measured at 77 K and 300 K, shown together with integral absorption spectra (dashed lines). The narrow line energies at a distance from the absorption edge are identical at different temperatures. Theoretical energies are marked at the bottom axes in (a) and (b).  (c) $\mu$--PL spectra measured in different crystals with $d\simeq$2.2 $\mu$m, exhibiting the similar energies of the WGM--related narrow lines.
\\
\\
{\bf Figure 5}

Temperature--induced mode switching in the cup-cavities. (a) SEM and $\mu$-CL images of the same InN crystal recorded at 5 K and 300 K showing the change of the dominant mode type from azimuthal to radial. (b) The simulated distributions of electromagnetic field intensity, as seen in the plane and cross--section of the crystal, shown together with the schematic shape used in these simulations. (c) Variation of the refractive index $n$ derived from the fitting of the WGMs at low (squares) and room (triangles) temperatures; the dotted lines is their polynomial fitting; the solid lines indicate the $n$ dependencies derived from the integral optical spectra. (d) Teraherz spectrum shown together with the peak of the IR emission comprising two unstable WGM--related lines. The energy scale is taken as $E-E_{0}$, where  $E_{0}$  is the peak energy of the first line for the IR emission, while for the THz band $E_{0}=0$.
\\
\\
{\bf Figure 6}

Dependencies of mode energy on the refractive index calculated for two neighboring modes in the InN cup--cavities. The simulated  spatial distributions of electromagnetic energy for these two modes are presented. The inset illustrates two possible regime: tuning and switching.

\end{document}